\PassOptionsToPackage{pdftex,
pdfencoding=auto,
pdfnewwindow=true,
pdfusetitle=true,
bookmarks=true,
bookmarksnumbered=true,
bookmarksopen=true,
pdfpagemode=UseThumbs,
bookmarksopenlevel=1,
pdfpagelabels=true,
breaklinks=true,
colorlinks=true,
}{hyperref}
\documentclass[amsmath,amssymb,superscriptaddress,twocolumn,notitlepage,nofootinbib, aps, pra, 10pt]{revtex4-2}

\usepackage[utf8]{inputenx}
\usepackage[english]{babel}
\usepackage{graphicx}
\usepackage{appendix}
\usepackage[dvipsnames]{xcolor}
\usepackage{soul}
\usepackage[]{hyperref}
\usepackage{microtype}
\microtypecontext{spacing=nonfrench}
\microtypesetup{
expansion={true,nocompatibility},
protrusion={true,nocompatibility},
activate={true,nocompatibility},
tracking=true,
kerning=true,
spacing={true}
}
\usepackage[all=normal,floats=tight,mathspacing=tight,wordspacing=tight,paragraphs=normal,tracking=tight,charwidths=tight]{savetrees}
\usepackage[style=american,autopunct=true]{csquotes} 
\usepackage{mathtools}
\usepackage{bbold}
\usepackage[flushleft, neveradjust]{paralist}
\usepackage{subfigure}
\usepackage{verbatim}
\usepackage{verbatimbox}
\usepackage{scalerel}
\usepackage[]{hyperref}
\hypersetup{linkcolor=carmine,citecolor=darkgreen,urlcolor=googleblue,anchorcolor=OliveGreen}

\definecolor{forestgreen}{rgb}{0.1, 0.35, 0.1}
\definecolor{ultramarine}{RGB}{63, 0, 255}
\definecolor{googleblue}{RGB}{34, 0, 204}
\definecolor{carmine}{RGB}{150, 0, 24}
\definecolor{darkgreen}{RGB}{0, 80, 0}

\newcommand{\ket}[1]{\left|#1\right\rangle}
\newcommand{\bra}[1]{\langle#1|}

\setstcolor{forestgreen}

\graphicspath{{./images/}}

\usepackage{tikz}
\usetikzlibrary{arrows.meta,fit,backgrounds}
\usepackage[skins,breakable]{tcolorbox}
\newtcolorbox{mdlp}[1]{enhanced,breakable,
  title={#1},fonttitle=\bfseries,coltitle=black,
  colbacktitle=black!8!white,
  colback=black!2!white,colframe=black!55!white,
  coltext=black,boxrule=.7pt,arc=1.2mm,
  borderline west={1.6pt}{0pt}{black!70!white},
  left=2mm,right=2mm,top=1.2mm,bottom=1.2mm,
  before upper={},
  before skip=8pt,after skip=8pt}
\definecolor{MDObservedText}{RGB}{1,32,96}
\definecolor{MDObservedEdge}{RGB}{47,82,143}
\definecolor{MDObservedFill}{RGB}{228,232,244}
\definecolor{MDLatentFill}{RGB}{251,229,205}
\definecolor{MDArrow}{RGB}{89,89,89}
\tikzset{
  mdgraph/.style={x=1cm,y=1cm,line width=.55pt,
    every node/.style={font=\fontsize{14}{16}\selectfont,text=black}},
  mdlatent/.style={draw=black,text=black,fill=MDLatentFill,
    rounded corners=2.5pt,inner sep=2pt,minimum size=8mm},
  mdobserved/.style={draw=MDObservedEdge,text=MDObservedText,
    fill=MDObservedFill,circle,inner sep=1.5pt,minimum size=8mm},
  mdcomposite/.style={draw=MDObservedEdge,fill=MDObservedFill,
    rounded corners=3pt,inner sep=6pt},
  mdcoordinate/.style={text=MDObservedText,inner sep=1.5pt},
  mdannotation/.style={font=\small,text=MDObservedText},
  mdarrow/.style={-{Triangle[length=2.1mm,width=1.6mm]},
    draw=MDArrow,line width=.9pt,shorten <=1pt,shorten >=1pt}
}
\newcommand{\mdBellPlain}{%
\begin{tikzpicture}[mdgraph]
\node[mdobserved] (x) at (-1.1,2.3) {$X$};
\node[mdobserved] (y) at (1.1,2.3) {$Y$};
\node[mdobserved] (a) at (-1.1,.6) {$A$};
\node[mdobserved] (b) at (1.1,.6) {$B$};
\node[mdlatent] (l) at (0,-.2) {$\Lambda$};
\draw[mdarrow] (x)--(a); \draw[mdarrow] (y)--(b);
\draw[mdarrow] (l)--(a); \draw[mdarrow] (l)--(b);
\end{tikzpicture}}
\newcommand{\mdBellDependent}{%
\begin{tikzpicture}[mdgraph]
\node[mdobserved] (x) at (-1.1,2.4) {$X$};
\node[mdobserved] (y) at (1.1,2.4) {$Y$};
\node[mdlatent] (l) at (0,1.2) {$\Lambda$};
\node[mdobserved] (a) at (-1.1,0) {$A$};
\node[mdobserved] (b) at (1.1,0) {$B$};
\draw[mdarrow] (l)--(x); \draw[mdarrow] (l)--(y);
\draw[mdarrow] (l)--(a); \draw[mdarrow] (l)--(b);
\draw[mdarrow] (x)--(a); \draw[mdarrow] (y)--(b);
\end{tikzpicture}}
\newcommand{\mdTriangle}{%
\begin{tikzpicture}[mdgraph]
\node[mdobserved] (c) at (0,2.8) {$C$};
\node[mdlatent] (g1) at (-1.2,1.85) {$\Gamma_1$};
\node[mdlatent] (g2) at (1.2,1.85) {$\Gamma_2$};
\node[mdobserved] (a) at (-1.2,.2) {$\tilde A$};
\node[mdobserved] (b) at (1.2,.2) {$\tilde B$};
\node[mdlatent] (l) at (0,-.7) {$\Lambda$};
\draw[mdarrow] (g1)--(c); \draw[mdarrow] (g2)--(c);
\draw[mdarrow] (g1)--(a); \draw[mdarrow] (g2)--(b);
\draw[mdarrow] (l)--(a); \draw[mdarrow] (l)--(b);
\end{tikzpicture}}
\newcommand{\mdFritz}{%
\begin{tikzpicture}[mdgraph]
\node[mdobserved] (c) at (0,2.9) {$C$};
\node[mdlatent] (g1) at (-2.35,1.75) {$\Gamma_1$};
\node[mdlatent] (g2) at (2.35,1.75) {$\Gamma_2$};
\node[mdobserved] (x) at (-1.1,1.05) {$X$};
\node[mdobserved] (y) at (1.1,1.05) {$Y$};
\node[mdlatent] (l) at (0,-.05) {$\Lambda$};
\node[mdobserved] (a) at (-1.1,-1.15) {$A$};
\node[mdobserved] (b) at (1.1,-1.15) {$B$};
\draw[mdarrow] (g1)--(c); \draw[mdarrow] (g2)--(c);
\draw[mdarrow] (g1)--(x); \draw[mdarrow] (g2)--(y);
\draw[mdarrow] (g1)--(a); \draw[mdarrow] (g2)--(b);
\draw[mdarrow] (l)--(x); \draw[mdarrow] (l)--(y);
\draw[mdarrow] (l)--(a); \draw[mdarrow] (l)--(b);
\draw[mdarrow] (x)--(a); \draw[mdarrow] (y)--(b);
\end{tikzpicture}}
\newcommand{\mdBellMarginal}{%
\begin{tikzpicture}[mdgraph]
\node[mdlatent] (g1) at (-2.35,1.75) {$\Gamma_1$};
\node[mdlatent] (g2) at (2.35,1.75) {$\Gamma_2$};
\node[mdobserved] (x) at (-1.1,1.05) {$X$};
\node[mdobserved] (y) at (1.1,1.05) {$Y$};
\node[mdlatent] (l) at (0,-.05) {$\Lambda$};
\node[mdobserved] (a) at (-1.1,-1.15) {$A$};
\node[mdobserved] (b) at (1.1,-1.15) {$B$};
\draw[mdarrow] (g1)--(x); \draw[mdarrow] (g2)--(y);
\draw[mdarrow] (g1)--(a); \draw[mdarrow] (g2)--(b);
\draw[mdarrow] (l)--(x); \draw[mdarrow] (l)--(y);
\draw[mdarrow] (l)--(a); \draw[mdarrow] (l)--(b);
\draw[mdarrow] (x)--(a); \draw[mdarrow] (y)--(b);
\end{tikzpicture}}
\newcommand{\mdUpperOriginal}{%
\begin{tikzpicture}[mdgraph]
\node[mdobserved] (c) at (0,2.8) {$C$};
\node[mdlatent] (g1) at (-1.6,1.7) {$\Gamma_1$};
\node[mdlatent] (g2) at (1.6,1.7) {$\Gamma_2$};
\node[mdobserved] (x) at (-1,.5) {$X$};
\node[mdobserved] (y) at (1,.5) {$Y$};
\node[mdlatent] (l) at (0,-.6) {$\Lambda$};
\draw[mdarrow] (g1)--(c); \draw[mdarrow] (g2)--(c);
\draw[mdarrow] (g1)--(x); \draw[mdarrow] (g2)--(y);
\draw[mdarrow] (l)--(x); \draw[mdarrow] (l)--(y);
\end{tikzpicture}}
\newcommand{\mdUpperMerged}{%
\begin{tikzpicture}[mdgraph]
\node[mdlatent] (g) at (0,2.8) {$\Gamma$};
\node[mdobserved] (x) at (-1.3,1.15) {$X$};
\node[mdobserved] (c) at (0,1.15) {$C$};
\node[mdobserved] (y) at (1.3,1.15) {$Y$};
\node[mdlatent] (l) at (0,-.6) {$\Lambda$};
\draw[mdarrow] (g)--(x); \draw[mdarrow] (g)--(c);
\draw[mdarrow] (g)--(y);
\draw[mdarrow] (l)--(x); \draw[mdarrow] (l)--(y);
\end{tikzpicture}}
\newcommand{\mdSettingBox}[1]{%
\node[mdcoordinate] (x) at (-.5,#1) {$X$};
\node[mdcoordinate] (y) at (.5,#1) {$Y$};
\begin{scope}[on background layer]
\node[mdcomposite,fit=(x)(y)] (z) {};
\end{scope}
\node[mdannotation,below=2pt] at (z.south) {$Z=(X,Y)$};}
\newcommand{\mdUpperCollider}{%
\begin{tikzpicture}[mdgraph]
\node[mdlatent] (l) at (-1.25,2.4) {$\Lambda$};
\node[mdobserved] (c) at (1.25,2.4) {$C$};
\mdSettingBox{.35}
\draw[mdarrow] (l)--(z.north west);
\draw[mdarrow] (c)--(z.north east);
\end{tikzpicture}}
\newcommand{\mdUpperColliderZ}{%
\begin{tikzpicture}[mdgraph]
\node[mdlatent] (l) at (-1.25,2.4) {$\Lambda$};
\node[mdobserved] (c) at (1.25,2.4) {$C$};
\node[mdobserved] (z) at (0,.35) {$Z$};
\draw[mdarrow] (l)--(z.north west);
\draw[mdarrow] (c)--(z.north east);
\end{tikzpicture}}
\newcommand{\mdUpperPrivate}{%
\begin{tikzpicture}[mdgraph]
\node[mdlatent] (l) at (-1.5,2.4) {$\Lambda$};
\node[mdlatent] (e) at (0,2.4) {$\epsilon_Z$};
\node[mdobserved] (c) at (1.5,2.4) {$C$};
\node[mdobserved] (z) at (0,.35) {$Z$};
\draw[mdarrow] (l)--(z.north west);
\draw[mdarrow] (e)--(z.north);
\draw[mdarrow] (c)--(z.north east);
\end{tikzpicture}}
\newcommand{\mdUpperFunction}{%
\begin{tikzpicture}[mdgraph]
\node[mdlatent] (l) at (-1,3.1) {$\Lambda$};
\node[mdlatent] (e) at (1,3.1) {$\epsilon_Z$};
\node[mdlatent] (t) at (0,1.7) {$\Lambda''$};
\node[mdobserved] (c) at (2.1,.35) {$C$};
\node[mdobserved] (z) at (0,.35) {$Z$};
\draw[mdarrow] (l)--(t); \draw[mdarrow] (e)--(t);
\draw[mdarrow] (t)--(z.north); \draw[mdarrow] (c)--(z.east);
\end{tikzpicture}}
\newcommand{\mdLowerChannel}{%
\begin{tikzpicture}[mdgraph]
\node[mdlatent] (l) at (0,2.4) {$\Lambda$};
\node[mdobserved] (x) at (-1.4,2.4) {$X$};
\node[mdobserved] (y) at (1.4,2.4) {$Y$};
\node[mdlatent] (t) at (0,1.1) {$\Lambda'$};
\node[mdobserved] (a) at (-1.4,0) {$A$};
\node[mdobserved] (b) at (1.4,0) {$B$};
\draw[mdarrow] (l)--(x); \draw[mdarrow] (l)--(y);
\draw[mdarrow] (l)--(t);
\draw[mdarrow] (t)--(a); \draw[mdarrow] (t)--(b);
\draw[mdarrow] (x)--(a); \draw[mdarrow] (y)--(b);
\end{tikzpicture}}

\begin{document}
%TC:ignore
\onecolumngrid
%\detailtexcount{main}
\twocolumngrid

\title{Bounding Freedom of Choice via Causal Network Embeddings}

\author{George Moreno}
\email{george.moreno@ufrn.br}
\affiliation{International Institute of Physics, Federal University of Rio Grande do Norte, 59078-970, P. O. Box 1613, Natal, Brazil}
\affiliation{Universidade Federal Rural de Pernambuco, 52171-900, Recife, Pernambuco, Brazil}

\author{Emanuele Polino}
\affiliation{Quantum and Advanced Technologies Research Institute, Griffith University, Yuggera Country, Brisbane, QLD 4111, Australia}

\author{Giovanni Rodari}
\affiliation{Dipartimento di Fisica - Sapienza Universit\`{a} di Roma, P.le Aldo Moro 5, I-00185 Roma, Italy}

\author{Davide Poderini}
\affiliation{Universit\`a degli Studi di Pavia, Dipartimento di Fisica, QUIT Group, via Bassi 6, 27100 Pavia, Italy}

\author{Askery Canabarro}
\affiliation{Quantum Research Center, Technology Innovation Institute, Abu Dhabi, UAE}
\affiliation{Grupo de F\'isica da Mat\'eria Condensada, N\'ucleo de Ci\^encias Exatas - NCEx, Campus Arapiraca, Universidade Federal de Alagoas, 57309-005, Arapiraca, AL, Brazil}
\affiliation{International Institute of Physics, Federal University of Rio Grande do Norte, 59078-970, P. O. Box 1613, Natal, Brazil}

\author{Elie Wolfe}
\affiliation{Perimeter Institute for Theoretical Physics, 31 Caroline St. N, Waterloo, Ontario, N2L 2Y5, Canada}

\author{Robert W. Spekkens}
\affiliation{Perimeter Institute for Theoretical Physics, 31 Caroline St. N, Waterloo, Ontario, N2L 2Y5, Canada}

\author{Fabio Sciarrino} 
\affiliation{Dipartimento di Fisica - Sapienza Universit\`{a} di Roma, P.le Aldo Moro 5, I-00185 Roma, Italy}

\author{Rafael Chaves}
\email{rafael.chaves@ufrn.br}
\affiliation{International Institute of Physics, Federal University of Rio Grande do Norte, 59078-970, P. O. Box 1613, Natal, Brazil}

\begin{abstract}
Standard Bell inequalities rely on measurement independence, the assumption that measurement settings are perfectly decoupled from the source of correlations. We demonstrate that embedding Bell scenarios into larger causal networks naturally overcomes this limitation. This embedding yields experimentally accessible lower and upper bounds on measurement dependence, the incompatibility of which certifies nonclassicality directly from data, without assuming a priori freedom of choice. We formulate a general linear-programming framework to systematically derive these measurement-dependence consistency (MDC) Bell inequalities. By bypassing standard non-convexity hurdles, this approach significantly outperforms existing techniques in computational simplicity and noise robustness. Finally, we reveal that genuine quantum nonlocality can be certified even for correlations exceeding the Tsirelson bound, and demonstrate our method's broad applicability to multi-outcome and multipartite Bell tests.
\end{abstract}
\maketitle

\section{Introduction}

Despite its foundational and practical significance, Bell's theorem \cite{bell1964einstein,Brunner} hinges on the seemingly untestable premise of measurement independence \cite{hall2016significance,handsteiner2017cosmic,rauch2018cosmic,big2018challenging}, the requirement that an observer's measurement choices remain entirely decoupled from the source generating the correlations. If this assumption is even mildly relaxed, classical causal models can perfectly mimic the violation of Bell inequalities, reaching up to their algebraic maximum~\cite{brans1988bell,hall2010local,barrett2011much,chaves2015unifying,putz2014arbitrarily,friedman2019relaxed,vieira2025test}. Consequently, if freedom of choice cannot be strictly guaranteed, standard Bell inequalities can no longer reliably certify nonclassicality, undermining their applications in quantum information processing.

Causal networks \cite{tavakoli2021bell} offer an alternative perspective on this limitation. By embedding a standard Bell experiment into a larger network with multiple independent sources, the burden of freedom of choice shifts from the measurement settings to the causal structure of the sources themselves~\cite{fritz2012beyond,chaves2021causal,polino2023experimental}. As Fritz demonstrated using a triangle network~\cite{fritz2012beyond}, assuming that sources are independent can indirectly enforce measurement independence at the level of observed correlations.

However, Fritz construction relies on perfect correlations between auxiliary variables, a condition that can never be matched in realistic experimental implementations. While recent approaches have addressed this limitation by attempting to bound measurement dependence directly from the additional network variables~\cite{fraser2018causal,chaves2021causal,polino2023experimental,lauand2025minimal}, they primarily rely on entropy-based inequalities \cite{chaves2014causal,fritz2012entropic,chaves2015information} or inflation techniques \cite{wolfe2019inflation,wolfe2021quantum}. These methods not only require computationally demanding, often non-convex optimization procedures but also struggle with noise robustness, yielding bounds that are not tight and too loose to be experimentally practical.

In this work, we overcome these limitations by developing a general framework to derive measurement-dependence consistency (MDC) inequalities in causal networks. Rather than treating measurement independence as an a priori assumption or assuming a fixed degree of dependence, we show that the freedom of choice compatible with the data can be bounded directly from observations. Specifically, we extract strict lower bounds from the embedded Bell subscenario alongside tight upper bounds constrained by the broader network structure. By formulating both problems as linear programs, this approach bypasses the non-convexity hurdles typically associated with causal networks. Ultimately, an incompatibility between these bounds rigorously certifies nonclassicality, providing tight and experimentally accessible constraints.

Applying this framework to noisy Fritz constructions, we demonstrate that nonclassicality can be certified without relying on perfect correlations. Our method significantly outperforms existing entropy- \cite{chaves2021causal} and inflation-based \cite{fraser2018causal} techniques, achieving superior noise robustness through much simpler optimization procedures. Furthermore, the framework remains effective beyond the quantum Tsirelson bound \cite{cirel1980quantum}, successfully distinguishing genuine quantum nonlocality from classical models that exploit measurement dependence to simulate post-quantum correlations \cite{popescu1994quantum}. To showcase its versatility, we extend our analysis beyond the standard Fritz and triangle networks to include embeddings of the Collins–Gisin–Linden–Massar–Popescu (CGLMP) scenario with multiple measurement outcomes \cite{Collins2002}, as well as multipartite extensions based on Mermin \cite{Mermin1990} and Svetlichny \cite{Svetlichny1987,Seevinck2002} inequalities. Ultimately, our results open new possibilities for device-independent tests of nonclassicality, providing rigorous certification in scenarios where traditional Bell inequalities are no longer sufficient.

The paper is organized as follows. In Sec.~\ref{sec:bellMI}, we review Bell's theorem and the role of measurement independence. Sec.~\ref{sec:triangle} discusses the triangle causal structure and how standard Bell scenarios can be embedded within it. In Sec.~\ref{sec:lps}, we introduce a general linear programming approach to bound measurement dependence and derive MDC inequalities. Sec.~\ref{sec:fritz} applies this framework to robustly witness the nonclassicality of the Fritz construction without assuming perfect correlations, extending the results to multi-party and multi-outcome scenarios. Sec.~\ref{sec:tsirelson} highlights an interesting consequence of measurement-dependent scenarios, the certification of quantum nonclassicality beyond Tsirelson's bound. Finally, Sec.~\ref{sec:discussion} summarizes our findings and outlines directions for future research.

\begin{figure*}[ht]
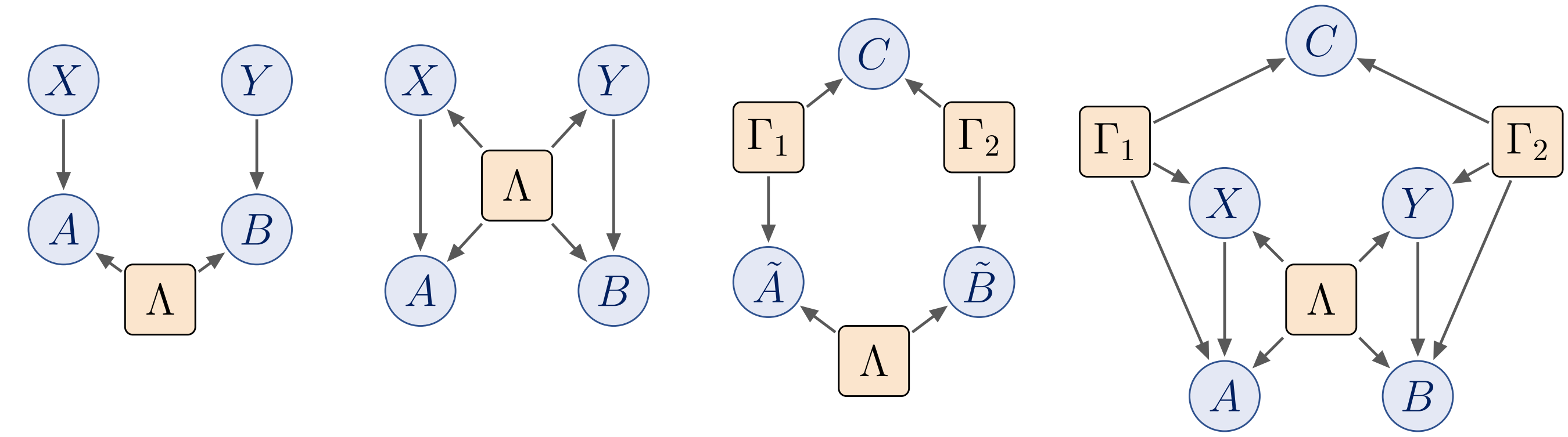

\centering
\subfigure[\textbf{Bell scenario.}]{%
\begin{minipage}[b]{.205\linewidth}\centering
\mdBellPlain\par\vspace{11mm}
\end{minipage}\label{fig: Bell_plain}}
\hfill
\subfigure[\textbf{Bell scenario with measurement dependence.}]{%
\begin{minipage}[b]{.205\linewidth}\centering
\mdBellDependent\par\vspace{12mm}
\end{minipage}\label{fig: Bell}}
\hfill
\subfigure[\textbf{Triangle scenario.}]{%
\begin{minipage}[b]{.205\linewidth}\centering
\mdTriangle\par\vspace{4mm}
\end{minipage}\label{fig: triangle}}
\hfill
\subfigure[\textbf{Bell scenario embedded in the triangle structure.}]{%
\begin{minipage}[b]{.32\linewidth}\centering
\mdFritz
\end{minipage}\label{fig: fritz_bell}}
\caption[]{ \textbf{Bell scenario embedded in the triangle scenario} \subref{fig: Bell_plain} Causal structure that represents the experiment proposed by Bell in his seminal paper \cite{bell1964einstein}.  \subref{fig: Bell}  Bell scenario featuring measurement dependence, i.e., causal dependences going from the latent variable $\lambda_{AB}$ towards the measurement choices $(x,y)$: such model is compatible with every no-signaling distribution $p(a,b|x,y)$. \subref{fig: triangle} Causal model of the triangle network. \subref{fig: fritz_bell} A embedding of the bipartite Bell scenario in the triangle DAG with $\tilde{A}=(A,X)$, $\tilde{B}=(B,Y)$ and $C=(C_0,C_1)$, all nodes being interconnected by three independent sources $\Gamma_1$, $\Gamma_2$ and $\Lambda$.}
\label{fig: freechoicedag}
\end{figure*}

\section{Bell's theorem and measurement independence}
\label{sec:bellMI}

The paradigmatic example of the incompatibility of quantum predictions with a classical causal model is given by Bell's theorem \cite{bell1964einstein}. In this scenario, one enforces an underlying causal structure represented by the directed acyclic graph (DAG) illustrated in Fig.~\ref{fig: Bell_plain}, which in a classical description implies that any observed probability distribution should allow for a local hidden variable (LHV) model given by
\begin{equation}
\label{eq:lhv}
p(a,b \vert x,y) = \sum_{\lambda} p(\lambda) p(a\vert x,\lambda)p(b\vert y, \lambda),
\end{equation}
where $X$ and $Y$ can be interpreted as the measurement choices of two distant observers, Alice and Bob, and $A$ and $B$ are the random variables representing the corresponding measurement outcomes. The absence of the directed edges from Alice's lab to Bob's lab implies that the statistics of measurement observed by Bob should be causally affected only by variables with direct causes to it, that is, $p(b\vert x,y,a,\lambda)=p(b\vert y,\lambda)$ and similarly $p(a\vert x,y,b,\lambda)=p(a\vert x,\lambda)$, also known as the local causality assumption. In turn, the absence of arrows between the source of correlations $\Lambda$ and the inputs $X$ and $Y$ implies that $p(x,y,\lambda)=p(x,y)p(\lambda)$, known as measurement independence, freedom of choice, or 'free-will' assumption. Classicality is then captured by assuming the existence of a shared classical variable $\Lambda$, governed by a well-defined probability distribution $p(\lambda)$, such that the conditional outcome distributions satisfy local causality.

The classical model in \eqref{eq:lhv} implies that any correlation compatible with it should respect constraints on the level of the observed probability distributions, constraints known as Bell inequalities. In the simplest Bell scenario, where each observer measures two possible dichotomic experiments ($x,y,a,b \in \{0,1\}$), the only non-trivial constraint is the Clauser-Horne-Shimony-Holt (CHSH) inequality~\cite{clauser1969proposed}
\begin{equation}
    \mathrm{CHSH} = \left|E_{00} + E_{01} + E_{10} - E_{11} \right| \leq 2 \, ,
    \label{eq:CHSH}
\end{equation}
where $E_{x,y} = \sum_{a,b} (-1)^{a + b} p(a,b \vert x, y)$.
Quantum theory, however, allows violations of this constraint up to $2\sqrt{2}$, known as Tsirelson's bound~\cite{cirel1980quantum}. Violations beyond the quantum bound are possible if one considers no-signalling (NS) correlations, those respecting locality at the observational level, that is, $p(a\vert x)=\sum_{b} p(a,b \vert x,y)=\sum_{b} p(a,b \vert x,y^{\prime})$ and $p(b\vert y)=\sum_{a} p(a,b \vert x,y)=\sum_{b} p(a,b \vert x^{\prime},y)$ and that can also be understood as those correlations respecting the LHV model in Eq. \eqref{eq:lhv}, however, with with a quasi-probability distribution $p(\lambda)$ \cite{eliens2020bell}. The paradigmatic example of a NS distribution being the Popescu-Rohrlich (PR)-box \cite{popescu1994quantum}, with a distribution
\begin{equation}
\label{eq:prbox}
p_{\mathrm{PR}}(a,b \vert x,y)= \frac{1}{2} \delta_{a\oplus b,xy},
\end{equation}
achieving $\mathrm{CHSH}=4$.

Notwithstanding, if the causal assumptions of locality and measurement independence are not enforced in the causal structure underlying an experiment, Bell inequality violations can be produced with classical causal models. In particular, if one allows for a measurement-dependent scenario, such as the one in Fig. \ref{fig: Bell}, even maximal NS violations of Bell inequalities can be mimicked. For example, assuming the DAG in Fig. \ref{fig: Bell} suppose that the source $\Lambda$ emits a triple of uniformly distributed classical bits $\lambda_1,\lambda_2,\lambda_3$, such that
\begin{eqnarray}
    p(\lambda_1,\lambda_2,\lambda_3)=\frac{1}{8} \quad \forall \lambda_1,\lambda_2,\lambda_3\;,
    \label{eq:triple}
\end{eqnarray}
and that the observed variables follow the instructions
\begin{align}
\label{eq: triple strategy}
    a &= \lambda_1,
    &
    b &= \lambda_1 \oplus \lambda_2\lambda_3\,,
    \\
    x &= \lambda_2\,,
    &
    y &= \lambda_3\;.
\end{align}
A direct calculation shows that this classical model reproduces the PR-box \eqref{eq:prbox} at the expense of introducing some measurement dependence, which can be quantified as \cite{chaves2015unifying}
\begin{equation}
    \mathcal{M} = \sum_{x,y,\lambda} \left| p(x, y,\lambda) - p(x, y)\, p(\lambda)\right| \, ,
\end{equation}
a trace distance quantifying how much the choices $X$ and $Y$ of the observers fail to be independent of the source $\Lambda$ and that for the particular model above gives $\mathcal{M} = 3/2$.

Given the violation of a certain Bell inequality, one can compute what is the minimum relaxation of measurement independence required to explain it with a classical causal model. In particular, for the CHSH scenario, one obtains \cite{chaves2015unifying}
\begin{equation}
    \label{eq:lower}
    \min\left(\mathcal{M}\right)
    =
    \max\left(0,\;\frac{I_{\mathrm{CHSH}} - 2}{4}\right),
\end{equation}
showing that larger CHSH violations necessarily require a larger departure from measurement independence. Remarkably, measurement dependence can act as a powerful simulation resource, as even small deviations from independence are sufficient to reproduce the full set of correlations achievable with maximally entangled states (see Refs. \cite{hall2010local,barrett2011much,putz2014arbitrarily} for a quantification using other measures of measurement dependence).

Notwithstanding, in a standard Bell scenario, the degree of measurement dependence is an a priori assumption of the underlying model and cannot be directly inferred from observed data alone. Consequently, measurement dependence remains an open and challenging loophole, motivating increasingly stringent experimental efforts to minimize its plausibility \cite{hall2016significance,handsteiner2017cosmic,rauch2018cosmic,big2018challenging}, although it cannot be completely ruled out.

\section{Embedding Bell's theorem in a triangle causal structure}
\label{sec:triangle}
The triangle network provides a paradigmatic example of a network generalization of Bell's theorem. Consider three parties, Alice, Bob, and Charlie, arranged in a triangle configuration such as in the DAG in Fig.\ref{fig: triangle}, where each pair of parties shares an independent source. The corresponding classical causal model assumes that the hidden variables associated with the three sources are mutually independent and that the observed correlations must therefore admit a decomposition of the form
\begin{equation}
\begin{aligned}
& &p(a,b,c)
=
\sum_{\lambda,\gamma_{1},\gamma_{2}}
p(\lambda)p(\gamma_{1})p(\gamma_{2})\\ \nonumber
& &\times p(a\vert\lambda,\gamma_{1})
p(b\vert\lambda,\gamma_{2})
p(c\vert\gamma_1,\gamma_2).
\end{aligned}
\end{equation}
Unlike the standard Bell-local decomposition, the independence of the three hidden variables imposes additional nonlinear constraints on the set of achievable correlations, giving rise to new forms of quantum non-classicality beyond those captured by Bell's theorem \cite{gisin2017elegant,renou2019genuine,suprano2022experimental,Sekatski2022partial,boreiri2025topologically,zamora2026latent,moradi2026fermions}. The first example of such non-classical behavior in the triangle network was proposed by Fritz \cite{fritz2012beyond}, who showed that a Bell scenario can be embedded within the triangle structure, allowing standard Bell nonlocality to be recovered as a special case of a more general network-based framework.

The core of Fritz's argument is to embed a standard Bell scenario inside the triangle network and invoke measurement independence. 
Formally, the original Bell setup is augmented by replacing $A$ and $B$ with $\tilde{A}=(A,X)$ and $\tilde{B}=(B,Y)$, respectively, and by introducing an additional node $C=(C_0,C_1)$, where $C_0$ and $C_1$ are binary variables. The node $C$ is linked through two independent latent sources, $\Gamma_1$ and $\Gamma_2$, as illustrated in Fig.~\ref{fig: fritz_bell}. Within this framework, measurement independence can be enforced by requiring perfect correlations between the pairs $(X,C_0)$ and $(Y,C_1)$. The argument is simple: if $X$ and $C_0$ are perfectly correlated, then any correlation between $X$ and $\Lambda$ would induce a correlation between $\Gamma_1$ and $\Lambda$, contradicting the assumption that these sources are independent. 
An analogous reasoning applies to $(Y,C_1)$, $\Gamma_2$, and $\Lambda$.

A quantum realization of such a scenario is given by
\begin{align}\label{qfritz}
p(a,b,x,y,c_0,c_1)
\\\nonumber =\mathrm{Tr}\big({\rho_{AB}} &\otimes {\rho_{X C_0}} \otimes {\rho_{Y C_1}}\,\cdot \,  {M^{A X}_{a,x}} \otimes {M^{B Y}_{b,y}} \otimes {M^{C_0 C_1}_{c_0,c_1}} \big)\;,
\end{align}
where $\rho_{A B}$ denotes the density operator of the state shared between Alice and Bob (thus replacing the classical description in terms of the hidden variable $\Lambda$) and similarly for $\rho_{X C_0}$ and $\rho_{Y C_1}$; $\{M^{A X}_{a,x}\}$ denotes a Positive Operator-Valued Measure (POVM) acting on the physical system in possession of Alice (similarly for $\{M^{B Y}_{b,y}\}$ and $\{M^{C_0 C_1}_{c_0,c_1}\}$). For the Fritz argument, the sources $\rho_{A B}$, $\rho_{X C_0}$, $\rho_{Y C_1}$ can be given by three maximally entangled states $\ket{\Phi}=(1/\sqrt{2})(\ket{00}+\ket{11}$). In turn, the POVMs are given by
\begin{align}\begin{split}\label{eq:FritzPOVMs}
& M^{C_0 C_1}_{(c_0,c_1)}=M^{C_0}_{c_0}\otimes M^{C_1}_{c_1}, \\
&
M^{A X}_{(a,x)}=M^{X}_{x}\otimes M^{A}_{a|x},
\\
&
M^{B Y}_{(b,y)}=M^{y}_{y}\otimes M^{B}_{b|y},
\end{split}
\end{align}
where $\{ M^{C_0}_{c_0}\},$ $\{M^{C_1}_{c_1}\},$ $\{M^{X}_{x}\},$ $\{M^{Y}_{y}\}$ are all 
measurements of the $\sigma_z$ basis, $ \{ M^{A}_{a|x}\}$ corresponds to one of the two Pauli observables among $\{\sigma_x, \sigma_z\}$, and $\{ M^{B}_{b|y} \}$ corresponds to one of the two observables among  $\{{(\sigma_z+\sigma_x)/\sqrt{2}}, {(\sigma_z-\sigma_x)/\sqrt{2}}\}$. The measurements are chosen to lead to perfect correlations between $X$ and $C_0$ and $Y$ and $C_1$, thus guaranteeing measurement independence; while at the same time ensuring that the conditional $p(a,b \vert x,y)$ violates a Bell inequality.

The Fritz argument, however, relies on perfect correlations that, in practice, will always be imperfect due to unavoidable noise. Consider, for example, that the source states are noisy and given by $\rho(v)=v\ket{\Phi}\bra{\Phi}+(1-v)\frac{\openone}{4}$. Then, $X$ and $C_0$, as well as $Y$ and $C_1$, are no longer perfectly correlated, preventing the direct application of the Fritz argument. Interestingly, although this noise does not introduce any measurement dependence, the resulting correlations are indistinguishable from those produced by a model where measurement independence is relaxed. In such more realistic scenarios, especially those where indeed some measurement dependence might be present, it requires taking into account the full causal structure of the network rather than relying solely on the correlations present in the data. Put in other terms, one should rely on Bell-like inequalities involving all variables in Fig.~\ref{fig: fritz_bell}.

Due to the intrinsic non-convexity of the set of correlations compatible with the triangle structure, obtaining a complete inequality description is, in general, highly non-trivial \cite{geiger1999quantifier,chaves2016polynomial}. This difficulty can be partially addressed using the inflation technique \cite{wolfe2019inflation}, as well as data-driven methods for deriving tailored causal compatibility inequalities \cite{fraser2018causal,polino2023experimental}. However, these approaches are typically sensitive to additional sources of noise beyond those for which the inequalities are specifically optimized, and they often require the evaluation of large and computationally demanding compatibility conditions.

An alternative strategy, closer in spirit to the original Fritz construction but not relying on perfect correlations, was proposed in Ref.~\cite{chaves2021causal}. The key idea is that, when embedding the problem in the triangle scenario, one can not only derive lower bounds on the measurement dependence parameter $\mathcal{M}$, such as in Eq.~\eqref{eq:lower}, but also obtain upper bounds by exploiting the additional observed variables $C_0$ and $C_1$, still under the assumption of a classical causal model. If the observed correlations admit a classical explanation, the lower bound must necessarily be no larger than the corresponding upper bound. That is, given a lower $\mathcal{M}_{\mathrm{lower}}$ and an upper $\mathcal{M}_{\mathrm{upper}}$ bound, it follows that
\begin{equation}
 \mathcal{M}_{\mathrm{lower}}\leq \mathcal{M}_{\mathrm{upper}},   
\end{equation}
a new family of classical causal constraints that we will generally refer to as measurement-dependence consistency (MDC) Bell inequalities. Conversely, a violation of this ordering certifies non-classicality. Despite their apparent simplicity, these inequalities are not only valid Bell-type constraints for the triangle scenario, but also constitute valid constraints in Bell scenarios allowing for measurement dependence. However, unlike in standard Bell analyses, where the degree of measurement dependence must be assumed a priori, in this framework, no such assumption is required. On the contrary, measurement dependence can, in principle, be arbitrary and can even be inferred or bounded from the observed data itself.

The issue with the approach taken in \cite{chaves2021causal} is that the obtained upper bounds rely on Shannon-entropy-based relaxations \cite{fritz2012entropic,chaves2014causal,chaves2014inferring}, which can be significantly loose. For instance, combining the lower- and upper-bound constructions yields \cite{chaves2021causal}
\begin{equation}\label{eq:multiMDexample}
\frac{\mathrm{CHSH}-2}{4}
\leq \sqrt{\frac{\Theta(X,Y,C_0,C_1)}{\log_2 e}},
\end{equation}
where $\Theta(X,Y,C_0,C_1)$ is a Shannon entropy–based quantity. Due to the poor tightness of the entropy bounds, violation of this inequality requires very high visibilities exceeding approximately $99.4\%$ for noisy Werner states $\rho(v)$.

\section{A general linear program approach to obtain Measurement-dependence consistency Bell inequalities}
\label{sec:lps}

The independence of the sources in the triangle scenario, as well as previous approaches based on the use of non-linear inequalities, such as those based on entropies or the inflation technique, seems to indicate that, unavoidably, the derivation of MDC Bell inequalities would require non-convex optimization tools. Fortunately, as we discuss next, that is not the case.

\begin{figure}[t!]
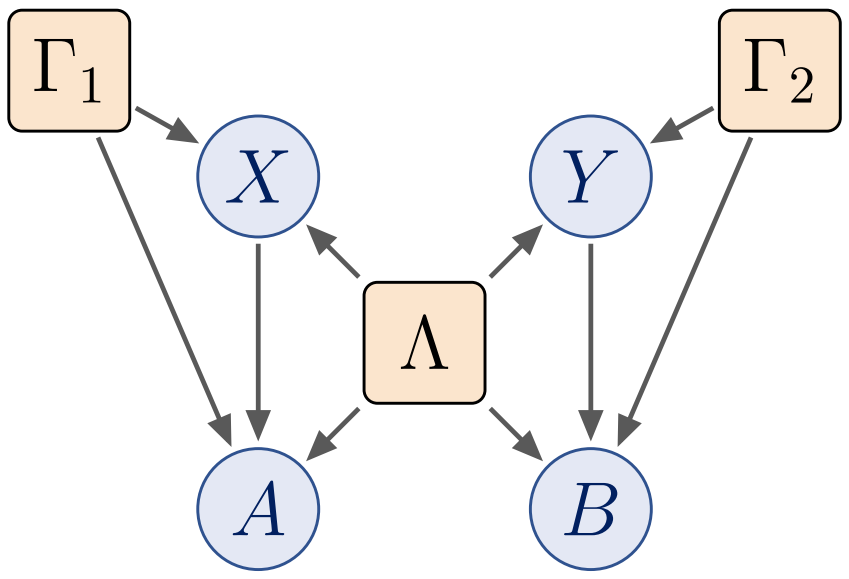

    \centering
    \mdBellMarginal
    \caption{\textbf{The subDAG from the \ref{fig: fritz_bell} representing a Bell scenario with measurement dependence}. This DAG is equivalent to the DAG in Fig.~\ref{fig: Bell} and allows us to lower bound the measurement dependence $\mathcal{M}_{X,Y:\Lambda}$.}
    \label{fig: Bell_part}
\end{figure}

To derive the lower bound $\mathcal{M}_{\mathrm{lower}}$, marginalize $C$ in Fig.~\ref{fig: fritz_bell}, obtaining Fig.~\ref{fig: Bell_part}. The resulting Bell marginal has local outcome responses conditioned on the source $\Lambda$, with measurement dependence allowed. Set $Z=(X,Y)$. A stochastic channel from $\Lambda$ samples a finite index $\Lambda'$ for a pair of deterministic local functions, without using the actual settings. This preserves $p(a,b,x,y)$ and obeys $Z-\Lambda-\Lambda'$. Data processing then gives $\mathcal M(Z:\Lambda')\leq\mathcal M(Z:\Lambda)$. Minimizing $\mathcal M(Z:\Lambda')$ over compatible joint distributions $q(z,\lambda')$ is a linear program~\cite{chaves2015unifying}. This optimization imposes no conditional independence between $X$ and $Y$ given $\Lambda'$. Appendix~\ref{sec:lower} gives the construction, the LP, and the resulting lower bound; see also the demo implementation at Ref.~\cite{code}.

\begin{figure}[t!]
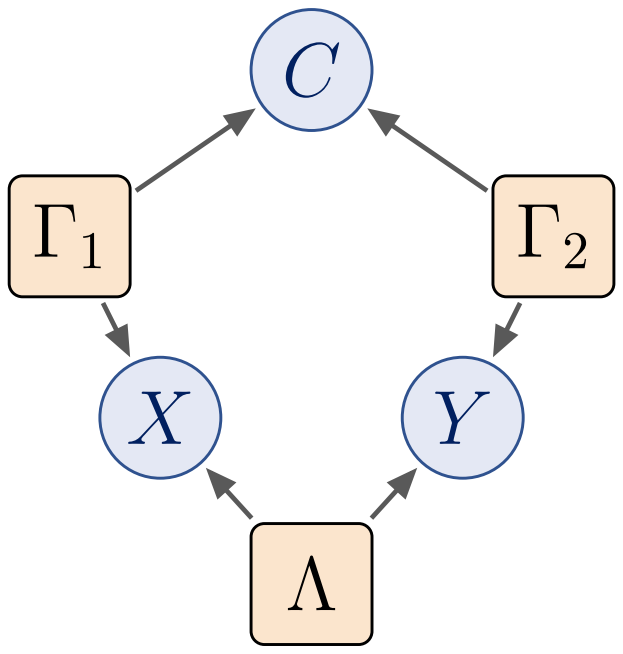

    \centering
    \mdUpperOriginal
 \caption{\textbf{A subDAG from the \ref{fig: fritz_bell} allowing to upper bound the measurement dependence $\mathcal{M}_{X,Y:\Lambda}$.}}
    \label{fig: upper_part}
\end{figure}

In turn, an upper bound $\mathcal{M}_{\mathrm{upper}}$ can be obtained by marginalizing over the nodes $A$ and $B$ in Fig.~\ref{fig: fritz_bell}. The resulting directed acyclic graph (DAG), shown in Fig.~\ref{fig: upper_part}, can again be recognized as an instance of the triangle scenario. In this setting, however, our goal is not to characterize which distributions $p(x,y,c)$ are compatible with the structure or exhibit nonclassical features, a case in which one certainly would require non-convex optimization tools. Instead, since our aim is rather to understand how the strength of the correlations between $(X,Y)$ and $C$ constrains the correlations between $(X,Y)$ and the latent variable $\Lambda$, one can convexify the problem and still obtain non-trivial upper bounds $\mathcal{M}_{\mathrm{upper}}$.

For each fixed original model, combine $\Gamma_1$ and $\Gamma_2$ as $\Gamma=(\Gamma_1,\Gamma_2)$ with their original product distribution. This preserves the original joint law; allowing arbitrary merged-source distributions and responses relaxes the model family while retaining $C\perp\Lambda$. With $Z=(X,Y)$, integrating out $\Gamma$ gives
\begin{equation}
q(z,c,\lambda)=q(\lambda)p(c)q(z\mid\lambda,c),
\label{eq: upper_factorization}
\end{equation}
where $p$ denotes observed distributions and $q$ denotes a posited model. The setting kernel is joint: $X$ and $Y$ need not be conditionally independent given $(\Lambda,C)$. Appendix~\ref{sec:upper} illustrates this representation and constructs a finite index $\Lambda''$ for deterministic functions $h_{\lambda''}:\mathcal C\to\mathcal Z$. For fixed $\Lambda$ and independent private randomness, $h_{\Lambda''}(c)$ specifies the joint kernel's outcome at each $c$. Evaluating $Z=h_{\Lambda''}(C)$ preserves the full original $q(z,c,\lambda)$ and gives the Markov chain $\Lambda-\Lambda''-Z$. Data processing through the reverse channel $q(\lambda\mid\lambda'')$ therefore gives
\begin{equation}
\mathcal M(Z:\Lambda)\leq\mathcal M(Z:\Lambda'')
\leq\mathcal M_{\mathrm{upper}}.
\label{eq: upper bound}
\end{equation}

For a deterministic function $h_{\lambda''}$, the conditional distribution is $q(z\mid\lambda'')=\sum_c p(c)\mathbb 1_{z=h_{\lambda''}(c)}$. Hence $\mathcal M(Z:\Lambda'')$ is linear in the weights $q(\lambda'')$. Maximizing it subject to reproducing the observed $p(z,c)$ gives the upper-bound LP in Appendix~\ref{sec:upper}. Only the artificial index $\Lambda''$ has a bounded alphabet: there are $|\mathcal Z|^{|\mathcal C|}$ deterministic functions (or $m^{2m_C}$ when $|\mathcal X|=|\mathcal Y|=m$ and $|\mathcal C|=m_C$). No cardinality bound on the original source $\Lambda$ is needed.

To illustrate the method, we consider two possible noisy implementations of the Fritz protocol. In the first, we consider that each of the sources is a noisy Bell state $\rho(\nu)$ with a visibility $\nu$ (NSS for noisy separated sources). In the second, we consider the ideal Fritz distribution mixed with a maximally mixed distribution, that is, $p(a,b,x,y,c_0,c_1)=\nu \, p_\mathrm{Fritz}(a,b,x,y,x,c_0,c_1)+ (1-\nu) 1/64$ (NF for noisy Fritz). In both cases, the quantum probability distribution $p(a,b \vert x,y)$, which will be employed to compute $\mathcal{M}_{\mathrm{lower}}$ via Eq. \eqref{eq:lower}, will be given by
\begin{eqnarray}
\label{eq: noisy_pr}
p_{\nu}(a,b|x,y)
=
\frac{\nu}{2\sqrt{2}}
\delta_{xy,a\oplus b}
+
\frac{\sqrt{2}-\nu}{4\sqrt{2}},
\end{eqnarray}
corresponding to a convex mixture of the quantum distribution attaining the maximal CHSH violation and uniform white noise with a corresponding CHSH value given $CHSH= 2\sqrt{2}\,\nu$.

In turn, for the distribution $p(x,y,c_0,c_1)$, which will be employed to compute $\mathcal{M}_{\mathrm{upper}}$, will be given by
\begin{eqnarray}
\label{eq: upper corr}
p(x,y,c_0,c_1)
=
\frac{\nu}{4}
\delta_{x,c_0}
\delta_{y,c_1}
+
(1-\nu)\frac{1}{16}
\end{eqnarray}
for the NF case, describing the perfect correlations obtained in the Fritz scenario mixed with white noise with visibility $\mu$. For the NSS scenario the distribution will be given by
\begin{eqnarray}
\label{eq: NS distribution}
\nonumber
    p(x,y,c_0,c_1) & = & \frac{\nu^2}{4}\delta_{x,c_0}
\delta_{y,c_1}
 + \frac{\nu(1-\nu)}{8}(\delta_{x,c_0} + \delta_{y,c_1})\\
 & + & \frac{1}{16}(1-\nu)^2
\end{eqnarray}
representing a noisy version of the sources between $(X,C_0)$ and $(Y,C_1)$, with a visibility parameter $\nu$.

In both cases, $\mathcal{M}_{\mathrm{upper}}$ can be computed through the corresponding linear program; see also the demo implementation at Ref.~\cite{code}. In the particular case of the NF model $\mathcal{M}_{\mathrm{upper}}$ scales linearly with $\nu$ according to
\begin{eqnarray}
\label{eq:upper}
 \mathcal{M}^{\mathrm{NF}}_{\mathrm{upper}} (\nu)
=
-\frac{3}{2}\nu+\frac{3}{2},
\end{eqnarray}
and taking $\nu=1$, we recover $\mathcal{M}_{\mathrm{upper}} = 0$, confirming the absence of measurement dependence. This aligns with Fritz’s original conclusion \cite{fritz2012beyond}, though there a different argument based on Shannon entropies was employed.

\section{A Fritz scenario without perfect correlations}
\label{sec:fritz}

The Fritz construction \cite{fritz2012beyond}, which embeds a Bell scenario into a triangle network, provided the first demonstration of certified nonclassical correlations in the triangle causal structure. A key ingredient of this argument, however, is the existence of perfect correlations between $X$ and $C_0$, and between $Y$ and $C_1$. Since such ideal correlations are unattainable in any realistic experiment, it is natural to ask whether nonclassicality can still be certified in the presence of noise. In this section, we show how our framework enables a robust implementation of a noisy version of the Fritz protocol.

\begin{figure}[t!]
    \centering
    \includegraphics[width=1.\linewidth]{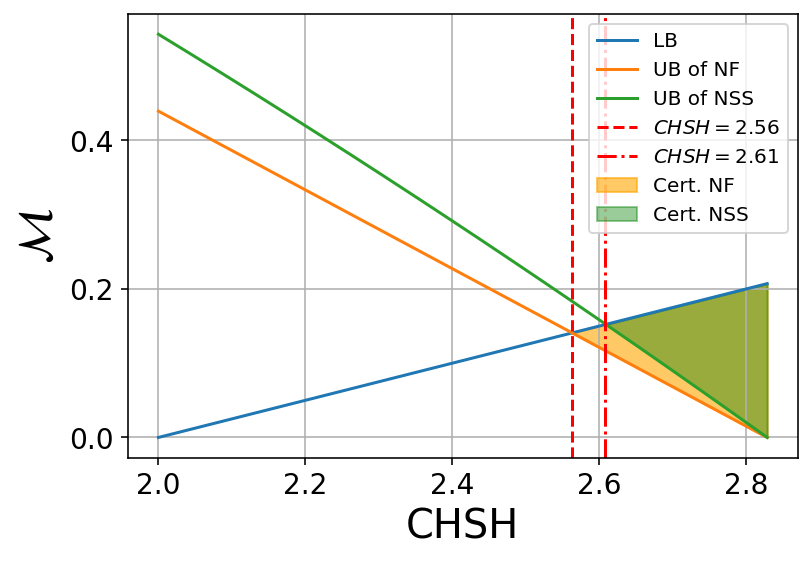}
    \caption{\textbf{MDC inequality violation for noisy Fritz distributions} Lower and upper bounds on the degree of measurement dependence. In green, the upper bound for the case NSS, sources of noisy Bell states $\rho(\nu)$ with a visibility $\nu$. In orange, the upper bound for the case NF, the ideal Fritz distribution mixed with the maximally mixed distribution. Nonclassicality is certified whenever the lower bound exceeds the upper bound (filled region).}
    \label{fig: md_plot}
\end{figure}

In the NF case, combining the lower bound in Eq.~\eqref{eq:lower} with the closed-form upper bound in Eq.~\eqref{eq:upper}, our method certifies non-classicality for any visibility above $\nu \geq 4(3-\sqrt{2})/7 \approx 0.906$ corresponding to $CHSH \approx 2.56$. In NSS case, we employ the distribution in Eq. \eqref{eq: NS distribution} to obtain non-classicality certification for visibilities above $\nu \approx 0.922$ corresponding to $CHSH \approx 2.61$. Both situations can be seen in Fig.~\ref{fig: md_plot}.

In the NF case, our method improves over the critical threshold visibility $\nu \approx 0.915$ obtained through large inflation graphs and the solution of much larger linear programs \cite{fraser2018causal}. In the NSS case, our method significantly improves over the threshold visibility obtained with the entropic bound and given by $\nu \approx 0.994$ \cite{chaves2021causal}. Our approach therefore achieves better certification power while remaining considerably simpler from both conceptual and computational perspectives.

\begin{figure}[t!]
    \centering
    \includegraphics[width=1.\linewidth]{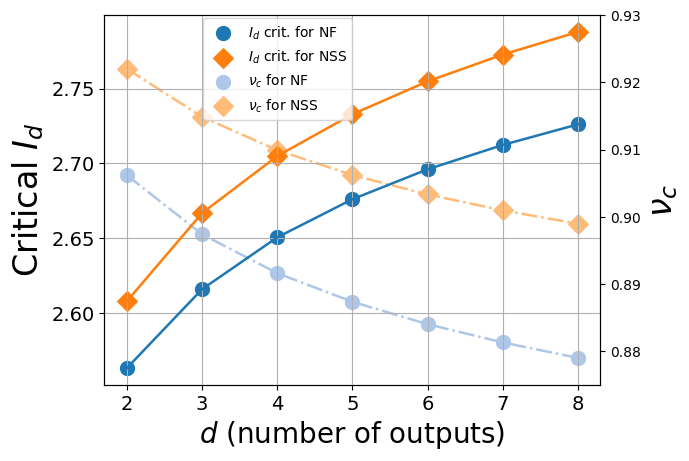}
    \caption{\textbf{Nonclassicality as a function of $d$ (measurement outputs)} In the left axis, the minimum critical value for the violation of $I_d$ inequality for the certification of nonclassicality via MDC inequalities. In the right axis, the critical visibility $\nu_c$ from which we can certify nonclassicality.}
    \label{fig: CGLMP}
\end{figure}

Furthermore, our approach is naturally generalizable to broader scenarios. For instance, we can explore the general result of Ref. \cite{chaves2015unifying} concerning the CGLMP inequality, stating that
\begin{eqnarray}
    \mathcal{M}_{\mbox{lower}} = \mbox{max} \left[0, \frac{I_d - 2}{4}\right]
\end{eqnarray}
in which $I_d\leq 2$ is the form of the CGLMP as displayed in \cite{Collins2002}. By extending the previous two-outcome scenario to a $d$-outcome experiment, we can leverage higher quantum bounds \cite{acin2002} to enhance noise robustness. Because the upper bound remains invariant (as the cardinalities of $C$, $X$, and $Y$ are fixed) and the maximum quantum violation of the $I_d$ inequality increases with $d$, the critical visibility required to violate the MDC inequality decreases with $d$. This is shown in Fig. \ref{fig: CGLMP}, considering that the visibility is added to the distribution as in the NF case discussed above. Curiously, in  spite of the increased robustness to the visibility, the minimum critical violation of the $I_d$ inequality increases with $d$.

Another possible generalization is to increase the number of parties in the Bell test. We approached this in two different ways, with the Svetlichny inequality \cite{Svetlichny1987,Seevinck2002}, of the form
\begin{eqnarray}
\nonumber
    S & = & E_{000} - E_{001} - E_{010} - E_{011}\\
    & - & E_{100} - E_{101} - E_{110} + E_{111}\leq 4,
\end{eqnarray}
and with the Mermin inequality \cite{Mermin1990}, of the form
\begin{eqnarray}
    M = E_{011} + E_{101} + E_{110} - E_{000} \leq 2.
\end{eqnarray}

Following our LP approach (see also the demo implementation at Ref.~\cite{code}), one obtains the lower and upper bounds
\begin{eqnarray}
    \mathcal{M}_{\mbox{lower}}^{S} = \mbox{max}\left[0, \frac{S - 4}{8}\right],
\end{eqnarray}
and
\begin{eqnarray}
    \mathcal{M}_{\mbox{lower}}^{M} = \mbox{max}\left[0, \frac{M - 2}{8}\right],
\end{eqnarray}
and
\begin{eqnarray}
\label{eq: tripartite upper}
    \mathcal{M}_{\mbox{upper}}= -\frac{7}{4}\nu + \frac{7}{4}.
\end{eqnarray}

Combining the bounds above, one can certify tripartite non-classicality without the need for measurement dependence for $M \geq 3.56$ corresponding to a critical visibility of $\nu \approx 0.89$ in the Mermin case, and $S \geq 5.18$ implying a critical visibility of $\nu\approx 0.92$ in the Svetlichny case. 

\section{Quantum nonclassicality beyond Tsirelson's bound}
\label{sec:tsirelson}

As we will discuss next, our approach allows certification of quantum nonclassical correlations even when the observed CHSH value exceeds Tsirelson's bound. In a standard bipartite Bell scenario such as the one in Fig. \ref{fig: Bell_plain}, any violation of the CHSH inequality beyond $2\sqrt{2}$ could only be attributed to a failure of the experiment to comply with the locality or measurement independence assumptions.

\begin{figure}
    \centering
    \includegraphics[width=1\linewidth]{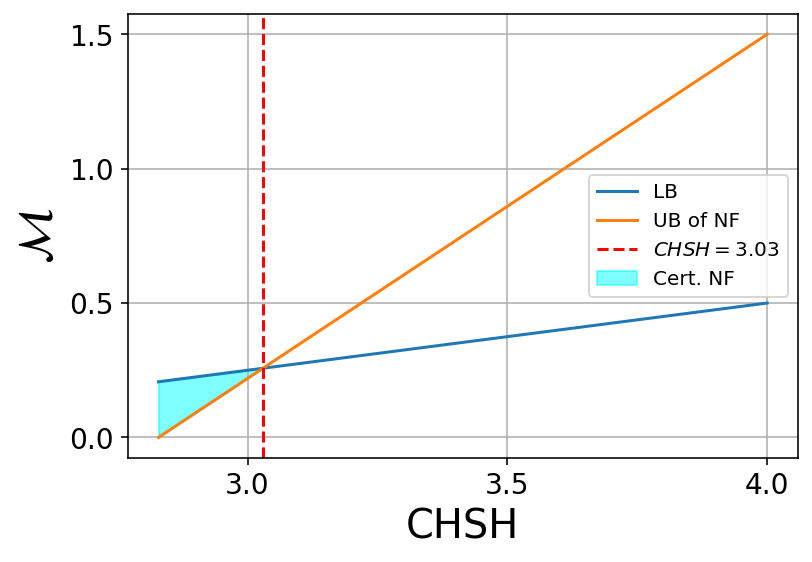}
    \caption{\textbf{Measurement dependence allows for quantum nonclassical correlation beyond the Tsirelson's bound} Highlight (in blue) of region for which nonlocality can be certified with the inconsistency between upper and lower bounds. The region lies in a strictly higher CHSH than the Tsirelson's bound, and indicates the capacity of distinguishing genuine nonlocality from a "fake nonlocality" yielded by an artificial PR-box.}
    \label{fig: old}
\end{figure}

Consider the triangle DAG as in Fig. \ref{fig: fritz_bell}, suppose now that the source $\Lambda$ operates according to an internal parameter $0\leq \nu \leq 1$. With probability $1-\nu$, this source employs the classical measurement-dependent strategy described in Eq. \eqref{eq:triple} which can be communicated to Alice and Bob via an ancillary flag variable. In turn, with probability $\nu$ the source $\Lambda$ distributes a maximally entangled state to Alice and Bob, which then perform measurements in the settings required to attain the maximal quantum violation of $2\sqrt{2}$ of the CHSH inequality. Let us suppose further that in this branch of the experiment, the variables $X$ and $Y$ are completely independent of $\Lambda$, that is, they are perfectly correlated with the node $C$ through the sources $\Gamma_1$ and $\Gamma_2$. The overall behavior, therefore, consists of a mixture of a genuinely quantum implementation and a classical simulation of PR-box correlations. The observed CHSH value is
\begin{equation}
    CHSH = \nu\,2\sqrt{2} + (1-\nu)\,4,
\end{equation}
which exceeds Tsirelson's bound whenever $\nu<1$. Thus, the lower bound on the degree of measurement dependence remains that of  Eq.~\eqref{eq:lower}. On the other hand, looking for the correlations among $C$, $X$ and $Y$, with probability $\nu$ they are perfectly correlated, while with probability $1-\nu$, we have a uniform distribution over their values. This implies that the upper bound for this experiment is exactly the same as in Eq. \eqref{eq:upper}. The obtained behavior is displayed in Fig.~\ref{fig: old}, showing that one can certify a quantum non-classical behaviour even beyond the Tsirelson's bound, up to a CHSH violation of $CHSH \approx 3.03$.

Crucially, the highlighted regions in Fig.~\ref{fig: old} can only be accessed by genuine nonclassical correlations. If the experiment relied entirely on classical simulations of post-quantum correlations, such as the measurement-dependent PR-box strategy, it could not simultaneously reproduce the observed CHSH value and the constrained measurement-dependence levels of this region. Therefore, the incompatibility between the upper and lower bounds serves as a definitive signature, distinguishing genuine quantum nonlocality from classical mechanisms designed to mimic stronger-than-quantum Bell violations.

\section{Discussion}
\label{sec:discussion}
In this work, we have developed a general framework to derive measurement-dependence consistency (MDC) inequalities in causal networks. Rather than treating measurement independence as an a priori assumption or assuming a fixed degree of dependence, we demonstrated that the freedom of choice compatible with the data can be bounded directly from observations. Specifically, we extracted strict lower bounds from the embedded Bell sub-scenario alongside tight upper bounds constrained by the broader network structure. By formulating both bounding problems as linear programs, this approach bypasses the non-convexity hurdles typically associated with causal networks. Ultimately, an incompatibility between these bounds rigorously certifies nonclassicality, providing tight and experimentally accessible constraints. 

Because our linear-programming approach systematically avoids these non-convexity hurdles, it significantly outperforms existing entropy- and inflation-based techniques. It offers superior noise robustness and computational simplicity when applied to realistic, noisy Fritz constructions, removing the need for perfect correlations. Furthermore, our framework remains effective even when observed correlations exceed the Tsirelson bound. By definitively ruling out classical models that exploit measurement dependence to mimic post-quantum correlations, our technique provides a sharp tool to distinguish genuine quantum nonlocality from classical simulation.  Beyond the standard bipartite setups, we have shown that our approach can readily be applied in multi-outcome as well as multipartite Bell scenarios.

Although our inequalities were explicitly derived with measurement-dependent Bell scenarios in mind, MDC inequalities constitute generally valid constraints for the broader network structure. Thus, a natural direction for future research involves investigating their potential as nonclassicality witnesses for wider families of distributions, extending beyond those strictly arising from measurement-dependent Bell tests. Furthermore, our framework could be adapted to quantify required levels of measurement dependence in other fundamental causal architectures, such as the bilocal network \cite{branciard2012bilocal,carvacho2017experimental,sun2019experimental}, a scenario where relaxations of freedom of choice remain largely unexplored. More broadly, the same approach could be used to investigate network nonclassicality in scenarios involving only separable measurements, encompassing the Fritz distribution \cite{fritz2012beyond} as a particular instance while extending naturally to a variety of other network configurations \cite{polino2026network}. Finally, synthesizing our methodology with the recently introduced concept of latent splitting, which has proven effective in certifying binary-outcome nonclassicality in the triangle network \cite{zamora2026latent} using an argument similar to that of Fritz \cite{fritz2012beyond}, offers another promising direction. Ultimately, these extensions could unlock even more powerful, device-independent tools for characterizing quantum correlations within complex network topologies.

\acknowledgements
We acknowledge financial support from the Simons Foundation (Grant No. 1023171, R.C.), the Brazilian National Council for Scientific and Technological Development (CNPq, Grants No. 168785/2023-4, 403181/2024-0, 301687/2025-0, 310992/2025-7), the National Institute of Science and Technology for Applied Quantum Computing through CNPq process No. 408884/2024-0, the Financiadora de Estudos e Projetos (Grant No. 1699/24 IIF-FINEP), the Fundação de Amparo à Ciência e Tecnologia do Estado de Pernambuco (Grant No. APQ-1702-1.05/24). We also thank the High-Performance Computing Center (NPAD) at UFRN for providing computational resources. E.P. acknowledges the support of the Australian Research Council through the Discovery Early Career Researcher Award grant DE250100762, and the Griffith University School of Environment and Science via the ESC Emerging Researcher Seed Grant.
The authors acknowledge support from FARE Ricerca in Italia QU-DICE Grant n. R20TRHTSPA and by the ERC Advanced Grant QU-BOSS (QUantum advantage via nonlinear BOSon Sampling, grant agreement no. 884676). 

The authors disclose that although the results in this manuscript were first proved without AI assistance, AI was subsequently used to simplify the proofs.
\bibliographystyle{ieeetr}
\bibliographystyle{naturemag}
\bibliography{mrefs}

\newpage
\begin{widetext}
\begin{appendices}
\renewcommand{\thesection}{\arabic{section}}
\renewcommand{\theHsection}{appendix.\arabic{section}}
\setcounter{equation}{0}
\renewcommand{\theequation}{S\arabic{equation}}
\renewcommand{\theHequation}{appendix.\arabic{equation}}
\makeatletter
\let\@hangfrom@section\@hangfrom@appendix
\let\@sectioncntformat\@appendixcntformat
\makeatother
\section{Obtaining a lower bound to measurement dependence}
\label{sec:lower}

The lower bound replaces the original source $\Lambda$ by a finite index $\Lambda'$ for deterministic local functions. Sampling this index without consulting the actual settings $Z=(X,Y)$ preserves $p(a,b,x,y)$ and gives $Z-\Lambda-\Lambda'$. Data processing can therefore only decrease measurement dependence; minimizing over compatible mixtures of these functions gives a lower bound.

Throughout these appendices, $p$ denotes observed distributions and $q$ a posited model. Observed alphabets are finite, but $\Lambda$ is unrestricted (replace sums by integrals when necessary); conditional kernels on null events are arbitrary. We use the full $\ell_1$ measure $\mathcal M(Z:H)=\sum_{z,h}|q(z,h)-p(z)q(h)|$ and write $\mathbb 1_E$ for the indicator of $E$.

\begin{figure*}[ht]
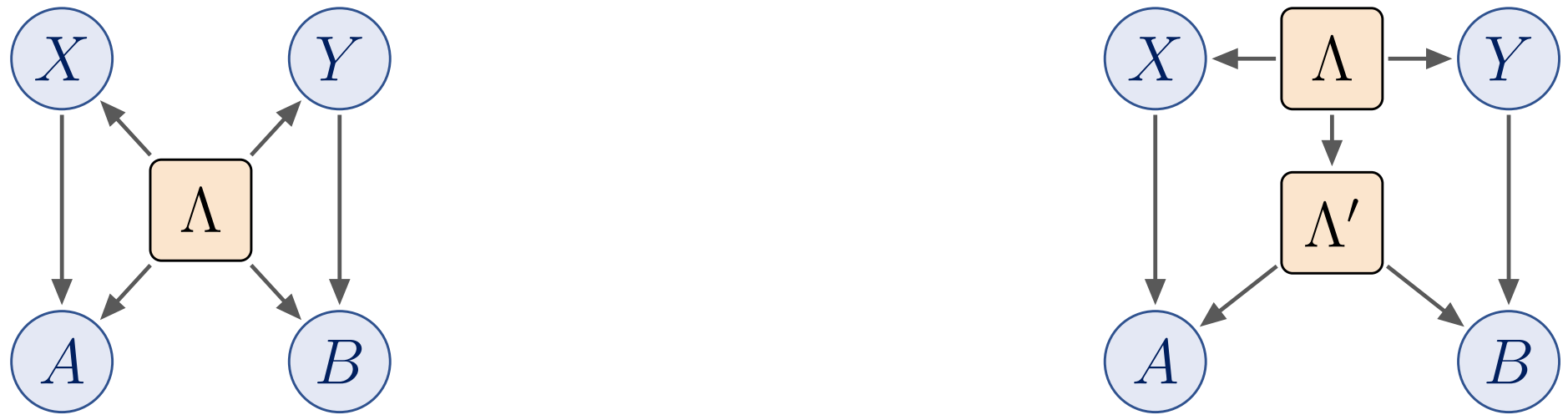

\centering
\begin{minipage}[b]{.44\linewidth}
\centering\mdBellDependent
\par\small (a) Original Bell model
\end{minipage}\hspace{.06\linewidth}%
\begin{minipage}[b]{.44\linewidth}
\centering\mdLowerChannel
\par\small (b) Sample and evaluate local functions
\end{minipage}
\caption{Lower-bound construction. The channel $\Lambda\to\Lambda'$ samples deterministic local functions before the actual settings determine $A=f_{\Lambda'}(X)$ and $B=g_{\Lambda'}(Y)$. This preserves both $q(z,\lambda)$ and $p(a,b,x,y)$.}
\label{fig:lower-construction}
\end{figure*}

\paragraph*{Deterministic local functions.}
Fix an original classical model. Its Bell marginal factors as $q(a,b,x,y,\lambda)=q(x,y,\lambda)q_A(a\mid x,\lambda)q_B(b\mid y,\lambda)$. For the triangle marginal, this follows by summing out $C$, using independence of $\Gamma_1,\Gamma_2$, and conditioning each local factor on its setting. Let $\lambda'$ index a pair of deterministic functions $f_{\lambda'}:\mathcal X\to\mathcal A$ and $g_{\lambda'}:\mathcal Y\to\mathcal B$. Sample this index via
\begin{equation}
K_L(\lambda'\mid\lambda)=
\prod_{u\in\mathcal X}q_A(f_{\lambda'}(u)\mid u,\lambda)
\prod_{v\in\mathcal Y}q_B(g_{\lambda'}(v)\mid v,\lambda),
\qquad q(z,\lambda,\lambda')=q(z,\lambda)K_L(\lambda'\mid\lambda).
\label{eq:lower-channel}
\end{equation}
The indices $u,v$ enumerate hypothetical settings; the channel never uses the actual $x,y$. Evaluating the functions at the actual settings and summing over their other values recovers the local response factors, hence $p(a,b,x,y)$. The Markov chain $Z-\Lambda-\Lambda'$ and the triangle inequality give
\begin{equation}
\mathcal M(Z:\Lambda')=
\sum_{z,\lambda'}\left|\sum_\lambda K_L(\lambda'\mid\lambda)
\bigl[q(z,\lambda)-p(z)q(\lambda)\bigr]\right|
\leq\mathcal M(Z:\Lambda).
\label{eq:lower-dpi}
\end{equation}

In the original triangle marginal $X\perp Y\mid\Lambda$, but mixing over $\Lambda$ need not leave $X\perp Y\mid\Lambda'$. For example, $X=Y=\Lambda$ with fair $\Lambda$ and constant outputs gives a constant index but correlated settings. We therefore allow arbitrary joint weights $q(x,y,\lambda')$, retaining $A=f_{\Lambda'}(X)$ and $B=g_{\Lambda'}(Y)$. Imposing setting independence would exclude valid constructions and introduce nonlinear constraints.

\begin{mdlp}{Lower-bound LP: deterministic functions from $X$ to $A$ and from $Y$ to $B$}
\textbf{Input:} $p(a,b,x,y)$ and all $|\mathcal A|^{|\mathcal X|}|\mathcal B|^{|\mathcal Y|}$ pairs $(f_{\lambda'},g_{\lambda'})$.\quad
\textbf{Variables:} joint weights $q(x,y,\lambda')\geq0$ and auxiliaries $t_{xy\lambda'}\geq0$, with $q(\lambda')=\sum_{x,y}q(x,y,\lambda')$.
\begin{equation}
\begin{aligned}
\mathcal M_{\mathrm{lower}}=\min_{q,t}\quad
&\sum_{x,y,\lambda'}t_{xy\lambda'}\\
\text{subject to}\quad
&\sum_{\lambda'}\mathbb 1_{a=f_{\lambda'}(x)}\mathbb 1_{b=g_{\lambda'}(y)}q(x,y,\lambda')=p(a,b,x,y),\\
&-t_{xy\lambda'}\leq q(x,y,\lambda')-p(x,y)q(\lambda')\leq t_{xy\lambda'}.
\end{aligned}
\label{eq:lower-lp}
\end{equation}
Constraints hold for every displayed index. The data constraints imply normalization and the correct setting marginal, including zero-probability settings. All coefficients are fixed; minimizing $t$ gives exactly $\mathcal M(Z:\Lambda')$~\cite{Boyd2004,chaves2015unifying}; see also the demo implementation at Ref.~\cite{code}.
\end{mdlp}

The constructed distribution is feasible, so every original classical model obeys
\begin{equation}
\boxed{\mathcal M_{\mathrm{lower}}\leq\mathcal M(Z:\Lambda')\leq\mathcal M(Z:\Lambda).}
\label{eq:lower-chain}
\end{equation}

\section{Obtaining an upper bound to measurement dependence}
\label{sec:upper}

The upper bound instead records more of the randomness generating $Z$. Since $C\perp\Lambda$, we can represent the joint response by $q(z\mid\lambda,c)$ and sample an index $\Lambda''$ for deterministic functions $h_{\lambda''}:\mathcal C\to\mathcal Z$. Evaluating $Z=h_{\Lambda''}(C)$ preserves the original $q(z,c,\lambda)$, while $\Lambda-\Lambda''-Z$ implies $\mathcal M(Z:\Lambda)\leq\mathcal M(Z:\Lambda'')$. Maximizing over compatible mixtures of these functions gives the upper bound.

\begin{figure*}[ht]
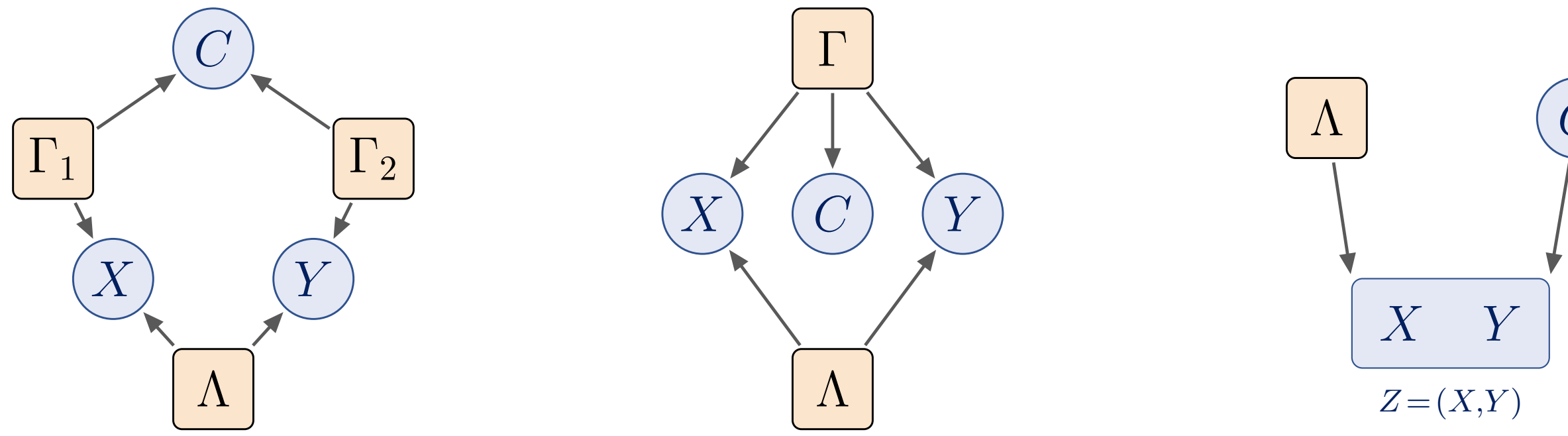

\centering
\begin{minipage}[b]{.31\linewidth}
\centering\mdUpperOriginal
\par\small (a) Original upper network
\end{minipage}\hfill
\begin{minipage}[b]{.31\linewidth}
\centering\mdUpperMerged
\par\small (b) Merge the upper sources
\end{minipage}\hfill
\begin{minipage}[b]{.31\linewidth}
\centering\mdUpperCollider
\par\small (c) Marginalize $\Gamma$
\end{minipage}
\caption{A joint setting kernel. Merging $\Gamma=(\Gamma_1,\Gamma_2)$ with its original distribution, then marginalizing it, preserves $q(z,c,\lambda)$. The box is one observed node $Z=(X,Y)$, allowing conditional correlations between its coordinates.}
\label{fig:upper-reduction}
\end{figure*}

\paragraph*{A joint setting kernel.}
For a fixed original model, merging $\Gamma=(\Gamma_1,\Gamma_2)$ with its product distribution and original response functions is only a relabeling. Allowing arbitrary merged-source distributions and dependence on the full merged source enlarges the model family. In either case, marginalizing $\Gamma$ yields
\begin{equation}
q(z,c,\lambda)=q(\lambda)p(c)q(z\mid\lambda,c).
\label{eq:upper-collider}
\end{equation}
For the fixed original model, this rewrite preserves the full $q(z,c,\lambda)$, hence $\mathcal M(Z:\Lambda)$. The setting kernel is joint: it need not factor into $q(x\mid\lambda,c)q(y\mid\lambda,c)$. Bounding all such kernels is sufficient.

\begin{figure*}[ht]
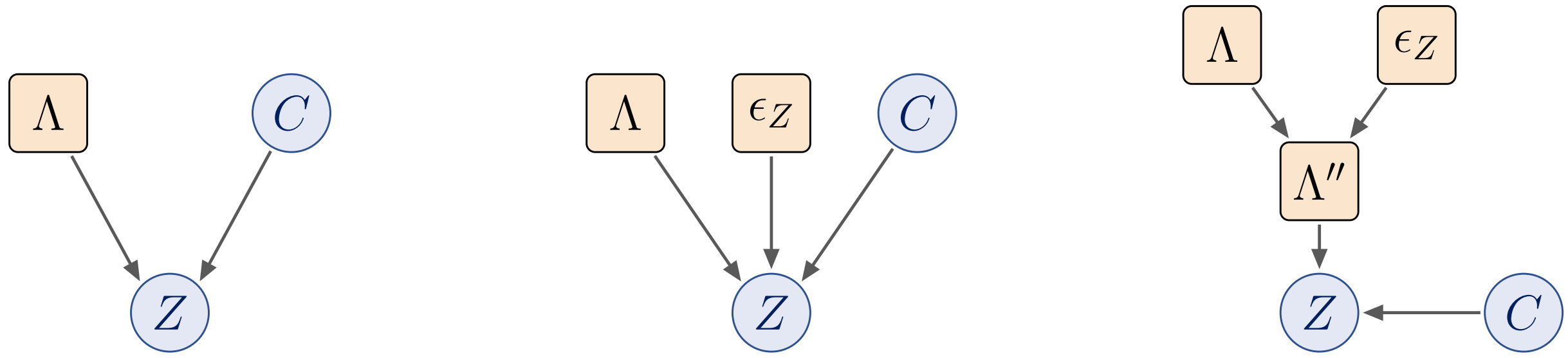

\centering
\begin{minipage}[b]{.31\linewidth}
\centering\mdUpperColliderZ
\par\small (a) The joint response kernel
\end{minipage}\hfill
\begin{minipage}[b]{.31\linewidth}
\centering\mdUpperPrivate
\par\small (b) Expose the response randomness
\end{minipage}\hfill
\begin{minipage}[b]{.31\linewidth}
\centering\mdUpperFunction
\par\small (c) Sample a deterministic function
\end{minipage}
\caption{Deterministic-function construction. Panel (a) represents Fig.~\ref{fig:upper-reduction}(c) using the single node $Z=(X,Y)$. The roots $\Lambda,C,\epsilon_Z$ are mutually independent. Sampling a deterministic function of $C$ preserves $q(z,c,\lambda)$; dependence on its index can exceed dependence on the original source.}
\label{fig:upper-construction}
\end{figure*}

\paragraph*{Deterministic joint response.}
Use private randomness $\epsilon_Z$, independent of $(\Lambda,C)$, to realize the collider response $Z=F(C,\Lambda,\epsilon_Z)$. Taking independent random numbers for the hypothetical values of $c$ gives a random deterministic function and the channel
\begin{equation}
h_{\Lambda''}(c)=F(c,\Lambda,\epsilon_Z)\quad(c\in\mathcal C),
\qquad K_U(\lambda''\mid\lambda)=\prod_{c\in\mathcal C}q(h_{\lambda''}(c)\mid\lambda,c),
\qquad Z=h_{\Lambda''}(C).
\label{eq:upper-function}
\end{equation}
Here $\Lambda''$ labels the sampled function, determined by $\Lambda$ and $\epsilon_Z$ independently of the actual $C$. The extension
\begin{equation}
q(z,c,\lambda,\lambda'')=q(\lambda)p(c)K_U(\lambda''\mid\lambda)\mathbb 1_{z=h_{\lambda''}(c)}
\label{eq:upper-extension}
\end{equation}
recovers Eq.~\eqref{eq:upper-collider} after summing over the function's other values. Thus the original joint law is preserved. Only the artificial index has a bounded alphabet, of size $|\mathcal Z|^{|\mathcal C|}$; no cardinality bound on $\Lambda$ is assumed. Since $C\perp(\Lambda,\Lambda'')$,
\begin{equation}
q(z\mid\lambda,\lambda'')=q(z\mid\lambda'')=
\sum_{c\in\mathcal C}p(c)\mathbb 1_{z=h_{\lambda''}(c)},
\qquad
q(z,\lambda)=\sum_{\lambda''}q(z,\lambda'')q(\lambda\mid\lambda'').
\label{eq:upper-reverse}
\end{equation}
This proves $\Lambda-\Lambda''-Z$: the reverse channel $q(\lambda\mid\lambda'')$ is independent of the actual $Z$. Applying the $\ell_1$ contraction in Eq.~\eqref{eq:lower-dpi} to this channel gives $\mathcal M(Z:\Lambda)\leq\mathcal M(Z:\Lambda'')$.

\begin{mdlp}{Upper-bound LP: deterministic functions from $C$ to $XY$}
Write $Z=(X,Y)$. \textbf{Input:} $p(z,c)$ and all $|\mathcal Z|^{|\mathcal C|}$ deterministic functions $h_{\lambda''}:\mathcal C\to\mathcal Z$, including constants.\quad
\textbf{Variables:} weights $q(\lambda'')\geq0$. Precompute the fixed coefficients
\begin{equation}
w_{\lambda''}=\sum_z\left|\sum_{c\in\mathcal C}p(c)\mathbb 1_{z=h_{\lambda''}(c)}-p(z)\right|.
\label{eq:upper-coefficient}
\end{equation}
Then $\mathcal M(Z:\Lambda'')=\sum_{\lambda''}q(\lambda'')w_{\lambda''}$, and
\begin{equation}
\begin{aligned}
\mathcal M_{\mathrm{upper}}=\max_q\quad
&\sum_{\lambda''}q(\lambda'')w_{\lambda''}\\
\text{subject to}\quad
&p(c)\sum_{\lambda''}q(\lambda'')\mathbb 1_{z=h_{\lambda''}(c)}=p(z,c)
\quad(z\in\mathcal Z,\ c\in\mathcal C),\\
&\sum_{\lambda''}q(\lambda'')=1.
\end{aligned}
\label{eq:upper-lp}
\end{equation}
The constraints use the full observed $p(z,c)$; only the objective coefficients use its marginals.\footnote{If $p(c)=0$, the corresponding data constraints are $0=0$ and that $c$ contributes nothing to the objective.} Every feasible solution realizes the collider relaxation by sampling $\Lambda''\perp C$ and evaluating $Z=h_{\Lambda''}(C)$.
\end{mdlp}

The constructed weights are feasible, giving
\begin{equation}
\boxed{\mathcal M(Z:\Lambda)\leq\mathcal M(Z:\Lambda'')\leq\mathcal M_{\mathrm{upper}}.}
\label{eq:upper-chain}
\end{equation}
The maximum is exact for the collider relaxation but need not be attainable in the original triangle network. Both constructions start from the same original $\Lambda$; their LP optimizers need not coincide with the constructed distributions. Combining the bounds gives
\begin{equation}
\boxed{\mathcal M_{\mathrm{lower}}\leq\mathcal M(Z:\Lambda')\leq\mathcal M(Z:\Lambda)\leq\mathcal M(Z:\Lambda'')\leq\mathcal M_{\mathrm{upper}}.}
\label{eq:five-term-chain}
\end{equation}
Thus $\mathcal M_{\mathrm{lower}}>\mathcal M_{\mathrm{upper}}$ rules out every original classical model; overlap is only necessary for compatibility. The constructions extend directly to more Bell parties by including all settings in $Z$ and all deterministic local functions in $\Lambda'$.
\end{appendices}
\end{widetext}

\end{document}